\documentclass[conference]{IEEEtran}
\IEEEoverridecommandlockouts
\usepackage{cite}
\usepackage{amsmath,amssymb,amsfonts}
\usepackage{graphicx, enumitem}
\usepackage{textcomp}
\usepackage[ruled,vlined,linesnumbered]{algorithm2e}
\SetAlFnt{\small}
\SetAlCapFnt{\small}
\SetAlCapNameFnt{\small}
\usepackage[table]{xcolor}
\usepackage{tikz}
\usepackage{bm}
\usepackage{booktabs}
\usepackage{multirow}
\usepackage{url}

\newcolumntype{L}[1]{>{\raggedright\let\newline\\\arraybackslash\hspace{0pt}}m{#1}}
\newcolumntype{C}[1]{>{\centering\let\newline\\\arraybackslash\hspace{0pt}}m{#1}}
\newcolumntype{R}[1]{>{\raggedleft\let\newline\\\arraybackslash\hspace{0pt}}m{#1}} 

\def\BibTeX{{\rm B\kern-.05em{\sc i\kern-.025em b}\kern-.08em
    T\kern-.1667em\lower.7ex\hbox{E}\kern-.125emX}}

\colorlet{pink}{red!40}
\colorlet{blue}{cyan!60}

\IEEEoverridecommandlockouts

\begin{document}

\title{Hybrid Quantum Computing - Tabu Search Algorithm for Partitioning Problems: preliminary study on the Traveling Salesman Problem}

\author{
\IEEEauthorblockN{Eneko Osaba\IEEEauthorrefmark{2}\IEEEauthorrefmark{1},
Esther Villar-Rodriguez\IEEEauthorrefmark{2}\IEEEauthorrefmark{1},
Izaskun Oregi\IEEEauthorrefmark{2},
and
Aitor Moreno-Fernandez-de-Leceta\IEEEauthorrefmark{3}}
\IEEEauthorblockA{\IEEEauthorrefmark{2}TECNALIA, Basque Research and Technology Alliance (BRTA), Mikeletegi Pasealekua 2, 20009, Donostia-San Sebastian, Spain}
\IEEEauthorblockA{\IEEEauthorrefmark{3} Instituto Ibermatica de Innovacion. Parque Tecnológico de Álava, Leonardo Da Vinci, 01510 Miñano, Spain\\
Email: [eneko.osaba, esther.villar, izaskun.oregui]@tecnalia.com, ai.moreno@ibermatica.com}
\IEEEauthorblockA{\IEEEauthorrefmark{1} Corresponding authors. These authors have equally contributed to the work presented in this paper.}}
\maketitle


\begin{abstract}
Quantum Computing is considered as the next frontier in computing, and it is attracting a lot of attention from the current scientific community. This kind of computation provides to researchers with a revolutionary paradigm for addressing complex optimization problems, offering a significant speed advantage and an efficient search ability. Anyway, Quantum Computing is still in an incipient stage of development. For this reason, present architectures show certain limitations, which have motivated the carrying out of this paper. In this paper, we introduce a novel solving scheme coined as hybrid \textit{Quantum Computing - Tabu Search} Algorithm. Main pillars of operation of the proposed method are a greater control over the access to quantum resources, and a considerable reduction of non-profitable accesses. To assess the quality of our method, we have used 7 different Traveling Salesman Problem instances as benchmarking set. The obtained outcomes support the preliminary conclusion that our algorithm is an approach which offers promising results for solving partitioning problems while it drastically reduces the access to quantum computing resources. We also contribute to the field of Transfer Optimization by developing an evolutionary multiform multitasking algorithm as initialization method.
\end{abstract}


\section{Introduction} \label{sec:intro}

Computational optimization is an intensively studied paradigm, which is the main focus of a vast number of works annually. The success of this knowledge field rests on the wide range of applications that it covers which yields a remarkable number of optimization types and problems. Furthermore, a myriad of techniques has been devised for efficiently solving such optimization problems. The competent tackling of optimization problems usually leads to the usage of notable computational resources. Even with the capacity of modern computers, the adoption of brute force methods is still impractical. Accordingly, the conception of alternative time-efficient solving schemes and paradigms is of paramount importance.

All these methods are conceived for their implementation and execution in classical computation devices. As an alternative to these classical systems, Quantum Computing (QC, \cite{steane1998quantum}) is emerging as a promising alternative for dealing with optimization problems. QC is considered as the next frontier in computation and it is attracting a lot of attention from the related community. QC offers to practitioners a revolutionary approach for addressing complex optimization problems, providing a significant speed advantage and an efficient search ability \cite{ajagekar2020quantum}. 

Today, two main types of QC architectures can be distinguished: the gate-model quantum computer and the annealing-based quantum computer \cite{wang201816,hauke2020perspectives}. In this study, we focus on the second alternative, which is the most applied technology in the current literature. Being more specific, among quantum-annealers, the most used commercial QC devices are the ones provided by D-Wave Systems \cite{gibney2017d}. Further details about QC architectures and D-Wave will be provided in upcoming sections.

In any case, although hopes placed in this field are very high and its potential advantages are really promising, QC is still in an incipient stage of development. Thus, current commercial QC architectures present huge limitations in terms of computational capabilities and performance \cite{fellous2020limitations,al2017natural}. Problems such as poor error correction, qubits prone to decoherence, and limited control of quantum resources entail important hindrances for rapid advances in this field.  

These limitations have motivated the conduction of this research. With this paper we take a step forward in this field by implementing a solving scheme for addressing partitioning problems. The main pillars of operation of our method are a greater control over the access to quantum resources and a considerable reduction of the non-profitable accesses. Our main motivation is that the efficient management of the the QC resources leads to a reduction in the economic costs. To reach these objectives, we present a hybrid \textit{Quantum Computing - Tabu Search} Algorithm, hereinafter referred to as Quantum Tabu Algorithm (QTA). Thus, our method contemplates three main stages for solving large partitioning problems: 1) the calculation of the problem partitions, 2) the efficient solving of the generated subproblems and 3) the merging of the independently-solved subinstances into a unique global solution. Thereupon, our method is specially conceived for addressing problems that can be solved by decomposition methods under the assumption of not trivial separability. The goal is to leverage the decomposition approach when the coupling amongst subproblems may undermine the performance of the primal problem. Moreover, without loss of generality, we have chosen the well-known Traveling Salesman Problem (TSP, \cite{applegate2006traveling}) with benchmarking purposes.

In a nutshell, this paper represents a step further over the current QC state of the art by elaborating on several research directions:

\begin{itemize}
    \item As far as we are concerned, most of the works proposed in the literature revolving around partitioning problems on QC devices are focused on the optimization of very small instances. This trend is a direct consequence of the limitations inherent to current QC computers and the problem related to the full access to these quantum platforms. In our work, we turn our attention to bigger problem instances composed by 14 to 23 nodes. Despite being small TSP instances in comparison to those in the classical computing state of the art, these sizes are much bigger than the instances often dealt by researchers belonging to quantum community.
    
    \item One of the major problems on QC research is the open access to quantum devices. Tackling big problem instances involves a partitioning because of the qubit restriction imposed by current QC machines. In this sense, D-Wave Systems offers alternatives such as QBSolv \cite{feld2019hybrid} or Hybrid Solvers. In any case, these alternatives also make a significant use to the QC device \cite{teplukhin2020electronic}, becoming prohibitive in terms of cost and prone to suffer any of the aforementioned limitations. Furthermore, configuration options provided by these algorithms are reduced, not allowing a fully control of the accesses to QC resources. Through the QTA, we reduce considerably the number of non-profitable calls made to the QC device, offering the user a fully configurable method for a supervised access to QC resources.
\end{itemize}

Last but not least, we also contribute to the field of Transfer Optimization (TO), \cite{gupta2017insights}. Specifically, we develop in this paper a multiform multitasking algorithm as initialization method for the developed QTA. Concretely, the algorithm implemented is a multiform variant of our previously published Coevolutionary Variable Neighborhood Search Algorithm (CoVNS, \cite{osaba2020CoVNS}). Further details are given in upcoming sections.

The rest of the paper is organized as follows: Section \ref{sec:back} presents a brief overview of the background related to this paper. In Section \ref{sec:QTA}, we describe the main characteristics of our proposed QTA. Experimental results are discussed in Section \ref{sec:exp}, along with a description of the benchmark and the experimental setup. Section \ref{sec:conc} concludes the paper by drawing conclusions and outlining future research lines.

\section{Background} \label{sec:back}

As stated in the introduction of this paper, this section is dedicated to providing a brief background on three main concepts studied in this paper: Quantum Computing (Section \ref{sec:quantum}), Transfer Optimization (Section \ref{sec:Transfer}), and the TSP (Section \ref{sec:TSP}).

\subsection{Quantum Computing} \label{sec:quantum}

Considered as the next frontier in the field of computation, QC surpasses all conventional computers in terms of power or performance, providing a significant speed advantage over classical methods \cite{nielsen2002quantum}. Technologically speaking, quantum computers are a specific class of devices which can conduct computation by leveraging quantum mechanical phenomena such as superposition and entanglement. 

These devices work with units of information called qubits \cite{hey1999quantum}, representing the state of a quantum particle. This quantum counterpart to the classical bit holds much more information than the binary digit by virtue of the superposition. This means that a qubit can be at both 1 and 0 states at the same time \cite{clark2019towards}. Thus, they can overcome the limitations of the classical binary representation. Different quantum states are possible for qubits until the measurement, when they collapse into one of their basis states \cite{ajagekar2020quantum}. Additionally, states may encode correlations among two or more qubits through entanglement. As a result, an action on a specific qubit will influence its entangled and connected qubits.

As mentioned in the introduction, two main types of QC architectures can be distinguished today: the gate-model quantum computer and the annealing-based quantum computer. The former is characterized by using quantum gates for the calculations and state manipulations of the quantum bits. The operation of these quantum gates can be compared to classical logic gates in traditional electronic circuits and they are sequentially applied to qubit states evolving up to the final solution of the problem at hand \cite{gyongyosi2018quantum}. Current commercial gate-model quantum computers have from 10 to 50 qubits in place, and some of their applications are integer factorization (Shor’s Algorithm) \cite{shor1999polynomial}, search algorithms (Grover’s Algorithm) \cite{grover1997quantum} or optimization problems (Quantum Approximate Optimization Algorithm) \cite{farhi2014quantum}.

On the other hand, the annealing-based quantum computers perform quantum annealing to find the state of minimum energy of a given quantum Hamiltonian (i.e., energy function). To this end, quantum-annealers make use of quantum phenomena such as tunneling and entanglement to reach lowest-energy state or ground-state. In this context, the Hamiltonian is programmed according to the objective function \cite{lucas2014ising} of the problem under study, so the lowest energy state represents the solution to the given problem. The leading provider of this kind of computers is D-Wave Systems. This company launched the \textit{D-Wave Advantage\_system1.1} computer, which is accessible via D-Wave’s cloud interface. This device has a working graph with 5436 qubits. In this computer, the Hamiltonian is expressed as an Isign model,
$\mathcal{H}_{\text{Ising}} = \sum_{i=1}^n h_i q_i + \sum_{j>i} J_{i,j} q_i q_j,$ where $q_i\in\{-1, 1\}$ represents the $i$-th qubit, $h_i$ is the linear bias associated to this variable, and $J_{i,j}$ represents the coupling strength  (the interaction) between $q_i$ and $q_j$ qubits. D-Wave allows for an alternative formulation to address optimization problems. This formulation corresponds to the Quadratic Unconstrained Binary Optimization (QUBO) problem \cite{glover2018tutorial}, which is mathematically expressed as:
\begin{equation}
    \mathbf{z}^* = \min_{\mathbf{z}\in\{0,1\}^{n}} \mathbf{z}^T \cdot \mathbf{Q} \cdot \mathbf{z}.
\end{equation} Here, $\mathbf{Q}$ is a programmable upper triangular matrix comprising the bias and couplers required by the Ising model. More precisely, the diagonal elements represent the linear biases and the nonzero off-diagonal terms the coupling coefficients. In this case, $\mathbf{z}\in\{0,1\}^{n}$ represents $n$-length binary variable array, and $\mathbf{z}^*$ the state that minimizes the quadratic function (i.e., the solution of the problem).

As can be easily seen in the literature, a myriad of works have been published in the field of QC using D-Wave device as their solver platform. Focusing our attention on studies related to the topic addressed in this paper, interesting research has been conducted on the Capacitated Vehicle Routing Problem (CVRP) \cite{feld2019hybrid}, Dynamic Multi-Depot CVRP \cite{harikrishnafkumar2020quantum}, the CVRP with time windows \cite{irie2019quantum} and different variants of the TSP \cite{kieu2019travelling}. The activity is significant and steady, hence illustrating perfectly the current condition of the community.

Turning our attention on the TSP, several remarkable works have been published in recent years proposing different schemes for problem formulation like \cite{warren2017small} or \cite{ruan2020quantum}. Among them all, we underscore the work proposed by Warren in 2020 in \cite{warren2020solving}, in which the author accurately describes the current state of the quantum community regarding the solving of the TSP. Furthermore, the author introduces some challenges and opportunities that should guide the subsequent studies.

Finally, it is convenient to finish this section bringing to the fore the QBSolv tool. The goal of this software is to help overcoming the infrastructure constraints when facing too large problems to be mapped directly to a QPU. Some examples of applications can be found in \cite{eagle2019solving} and \cite{okada2019improving}. In a nutshell, QBSolv is a tool provided by D-Wave which splits the complete QUBO into smaller subQUBOs and solves them sequentially and independently. This process is iteratively run until no improvement is achieved in the problem solution. Going deeper, QBSolv breaks the complete QUBO matrix through the use of a tabu search heuristic, and it can be used both locally and on the D-Wave hardware as a quantum-classic hybrid approach. We recommend \cite{booth2017partitioning} for additional details about QBSolv.

Anyway, hybrid alternatives such as QBSolv have their own limitations, as can be read in recent works such as \cite{teplukhin2020electronic}. For example, the application of this strategy implies a massive use of the QC device. Furthermore, although QBSolv is slightly configurable, a proper setting of its parameters also demands a high number of calls to the quantum service. This fact hampers the D-Wave hardware access monitoring and management. 

Bearing this background in mind, this work represents a step further over the current QC state of the art by elaborating on the directions pointed out in the introduction of this paper.

\subsection{Transfer Optimization - Evolutionary Multitasking}\label{sec:Transfer}

Transfer Optimization is a relatively new knowledge field within the wider area of optimization. The core idea lies in the exploitation of prior knowledge, learnt from the optimization of one problem, to get substantial benefits when facing another related or unrelated problem (or tasks) \cite{gupta2017insights}. This concept comprises three solving approaches: \textit{sequential transfer} in which a target task gets advantage from a knowledge base encompassing previous task information; \textit{multitasking} which refers to the simultaneous solving of different problems of equal importance by dynamically exploiting existing synergies; and the so-called \textit{multiform optimization} where a single task is solved using different problem formulations, being all of them optimized simultaneously through a multitasking perspective.

Specifically, we adopt the paradigm known as Evolutionary Multitasking (EM, \cite{ong2016towards}) which embraces the concepts, search strategies and operators conceived within Evolutionary Computation for simultaneously dealing with several tasks. Through the design of a unified search space, these population-based algorithms permit the parallel evolution of the whole set of problems to optimize, allowing also the constant sharing of genetic material among solutions aiming to exploit inter-task synergies.

Going deeper, a consensus was reached on multifactorial optimization as the only formulation of EM until 2017 \cite{gupta2015multifactorial}. From that moment on, a remarkable amount of works have been published in the literature proposing interesting alternative algorithmic schemes such as the multitasking multi-swarm optimization proposed in \cite{song2019multitasking} or the coevolutionary bat algorithm introduced in \cite{osaba2020coeba} among others. Regarding Multifactorial Optimization, apart from the well-known Multifactorial Evolutionary Algorithm \cite{gupta2015multifactorial}, some additional methods have been recently developed by the related community. Valuable examples are the particle swarm optimization-firefly hybridization defined in \cite{xiao2019multifactorial} or the multifactorial cellular genetic algorithm implemented in \cite{osaba2020multifactorial}.

As mentioned in the introduction, we make a significant contribution to this specific research area by adapting our recently published CoVNS \cite{osaba2020CoVNS} not only to this specific application, but also to the multiform paradigm, which is scarcely studied in the related literature.

\subsection{Traveling Salesman Problem as Benchmarking Problem}\label{sec:TSP}

As mentioned in the introduction of this paper, this work introduces a preliminary study employing the TSP as benchmarking problem. Formally, the TSP can be defined as a complete graph $G=(V,A)$, where $V= \{v_1,v_2,\dots,v_N\}$ is the set of $N=|V|$ vertices of the system, and $A= \{(v_i,v_j): (v_i,v_j)  \in V^2 \text{ for }i\neq j \}$ the group of arcs interconnecting these nodes. Besides that, each arc has an associated $c_{ij}$ cost, which is the same in both directions, i.e., $c_{ij}$ = $c_{ji}$. Thus, the objective of TSP is to calculate a complete $TSP^*$ that, starting and finishing at the same point, visits every node once and minimizes the total cost of the path.

As widely known, the TSP is one of the most intensively studied optimization problems since it still poses a challenge as a NP-hard problem and due to its straightforward adaptation to real-world problems \cite{osaba2020traveling}. Thence, a plethora of intelligent methods have been applied along the last decades. Nevertheless, advanced and sophisticated bio-inspired methods have lately flourished and reached enhanced performance. The principal goal of this paper is not to optimally solve the TSP, but to statistically and fairly compare the performance of the considered solvers under the same instances and conditions.

It is crucial to spotlight in this section that the QUBO formulation used in this paper for the TSP is based on the one proposed by Feld et al. in \cite{feld2019hybrid}. Due to extension restrictions, we recommend the analysis of that paper to get additional details on this QUBO formulation.

\section{Hybrid Quantum Computing - Tabu Search Algorithm}\label{sec:QTA}

\begin{figure}[t]
	\centering
	\includegraphics[width=0.9\hsize]{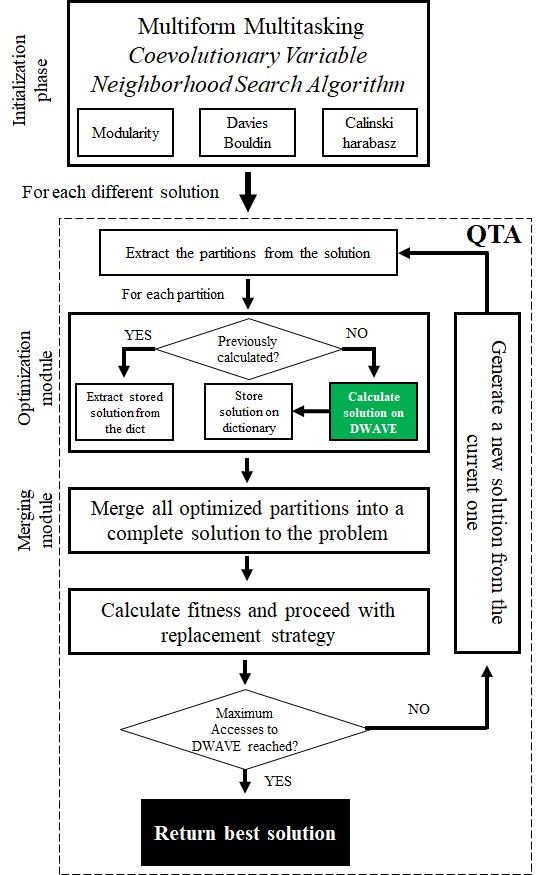}
	\caption{Workflow of QTA including initialization procedure used in this paper.}
	\label{fig:QTA}
\end{figure}

The proposed QTA is composed by different modules, which have been depicted in Figure \ref{fig:QTA}. In order to properly describe the designed and implemented approach, we devote different subsections here to outline the most important aspects of the QTA.

\subsection{Initialization procedure}\label{sec:init}

This section describes the initialization phase as a novel contribution to the TO research field. If researchers opt not to use any initialization function, we highly recommend starting the execution of QTA with randomly generated feasible solutions.

Specifically, the initialization algorithm is an adaptation of the aforementioned CoVNS. Briefly explained, CoVNS is a coevolutionary multitasking algorithm which is composed by different subpopulations of individuals. Each subpopulation is devoted to the optimization of a single task through the execution of a VNS. Furthermore, the coevolution is materialized through the periodical sharing of individuals which migrate amongst subpopulations in the hope of contributing to a faster convergence in the related tasks. On the grounds that an optimum graph partition does not necessarily entail a good decomposition scheme for the problem at hand in pursuit for an optimum ulterior recomposition, we leverage this multiform algorithm to explore distinct clustering solutions by means of 3 extensively-applied metrics: Modularity \cite{newman2004finding}, Davies Boulding \cite{davies1979cluster} and Calinski Harabasz \cite{calinski1974dendrite}. Thus, and following the procedure deeply detailed in \cite{osaba2020CoVNS}, the multiform multitasking CoVNS provides the best clustering partition found by each subpopulation according to a particular metric.

\subsection{Optimization Module}\label{sec:opt}

Before starting with the description of the optimization module, it is worthy explaining the codification used in this paper for representing the clustering proposed for each TSP instance. Specifically, we have adopted the representation known as label-based \cite{hruschka2009survey}, in which a solution is encoded using a $N$-dimensional vector $\mathbf{x}=[x_1,\ldots,x_n,\ldots, x_N]$ where $x_n\in\{1,\ldots,C\}$ represents the cluster label the $n$-th node belongs to and $N$ denotes the total number of nodes in the network, i.e. the number of stops of the TSP instance. Thus, such clusters give birth to a $C$ group of subgraphs $\widetilde{\mathcal{G}}$, being $\widetilde{\mathcal{G}^
i}$ (partition for $i$-th cluster) to be solved as an independent $TSP^i$. For instance, assuming a TSP instance with $N=9$ nodes, $\mathbf{x}=[1,1,2,1,3,3,2,3,2]$ represents one possible feasible solution with $C = 3$ different partitions $\widetilde{\mathcal{G}}=\{\mathcal{G}^1,\mathcal{G}^2,\mathcal{G}^3\}$, where $\mathcal{G}^1=\{1,2,4\}$, $\mathcal{G}^2=\{3,7,9\}$ and $\mathcal{G}^3=\{5,6,8\}$.

Developed QTA starts its execution by taking the initial solution (either initialized or randomly generated) and extracting all partitions/clusters $\widetilde{\mathcal{G}}$ present in that individual. Then, considering each partition as a subproblem, QTA calculates the route $TSP^i$ for each of these clusters $\widetilde{\mathcal{G}^i}$. At this point, it is compulsory to specify that the optimization of each partition is conducted through the D-Wave Quantum Annealer. In this sense, and with the principal objective of minimizing the remote accesses to this quantum computer, QTA stores all calculated subroutes in a resource coined as \textit{Tabu Dictionary}. Thus, this dictionary collects all previously visited stages. In this way, once the algorithm receives a subproblem to be optimized, QTA checks if this particular partition has been previously treated. In positive case, QTA obtains the previously calculated solution, avoiding the access to the remote D-Wave. If negative, the subproblem is sent to the D-Wave Quantum Annealer storing the computed solution in the \textit{Tabu Dictionary} for further use. Basically, the main motivation behind the use of this \textit{Tabu Dictionary} mechanism is to avoid repeated and unnecessary accesses to the QC device. 

\begin{figure}[t]
	\centering
	\includegraphics[width=0.95\hsize]{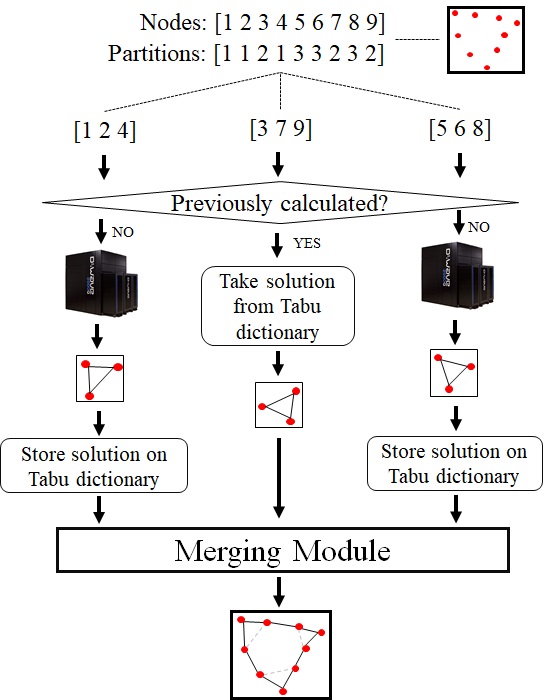}
	\caption{Workflow of the Optimization and Merging modules taking as example a 9-node TSP instance.}
	\label{fig:optimization}
\end{figure}

We illustrate the whole procedure in Figure \ref{fig:optimization} by taking the TSP example explained above as use case. The figure also covers the adjoining Merging Module, right after detailed.

\subsection{Merging Module}\label{sec:merge}

The merging module follows a greedy strategy in the search of composing the complete solution $TSP^*$ across the pool of $\widetilde{\mathcal{TSP}}$ subroutes. This $TSP^*$ comprises the arcs belonging to every $TSP^{i}$ except those arcs, one for each $TSP^{i}$, required to open the closed loops-subroutes. The composition is then performed by the creation of $C$ bridges, each one responsible for the linkage of two partitions in $\widetilde{\mathcal{G}}$ through two disconnected nodes.  

At the beginning, Algorithm \ref{alg:merging} in charge of conducting this merging starts from a random subroute $TSP^{i}$ in cluster $i$, by randomly selecting an internal arc $ia=(v_{m}^{i}, v_{n}^{i})$. Concretely, one internal arc in $TSP^{i}$ must be removed to enable the creation of a bridge between subgraph ${\mathcal{G}^i}$ and another ${\mathcal{G}^{new}}$, being initially $v_{m}^{i}$ the potential connector to an adjoined cluster and $v_{n}^{i}$ the orphan node awaiting the last created bridge that closes the complete Hamiltonian cycle. Then, Algorithm \ref{alg:GetNextLink} is called, firstly taking $ia$ (internal broken edge in $TSP_{i}$), and node $v_{m}^{i}$ is linked to the closest external node $v_{s}^{new}$ creating a new inter-cluster bridge $ea=(v_{m}^{i}, v_{s}^{new})$ towards a non-visited cluster $G^{new}$. $v_{s}^{new}$ will provide two opposite directions to walk through in $TSP^{new}$, both to be explored. This composition procedure \{break internal arc - create bridge\} is repeated by means of Algorithm \ref{alg:RecursiveSearch} in a recursive fashion, until $v_{n}^{i}$ is reached. 

Due to the greedy nature of this search strategy, the arbitrary subroute $TSP^{i}$ to start from and the two potential intra-cluster arcs to break in every $TSP^{new}$ yield a collection of potentially minimum routes in terms of total distance stored in an agglomerating tree structure $cycles$. The best found path will be returned as depicted in Algorithm \ref{alg:merging}, once minimal distance for every root-to-leaf path is estimated.

\begin{algorithm}[h!]
    \SetKwInOut{Input}{input}
    \SetKwInOut{Output}{output}
    
	\Input{$\widetilde{\mathcal{TSP}}=\{TSP^i\}_{i=1}^{C}$}
	Initialize: $linkedClusters$, $cycles$\;
	\For{$TSP^i \in \widetilde{\mathcal{TSP}}$}{
	    Insert($i$) into $linkedClusters$\;
	    \For{$ia(v_{m}^{i}, v_{n}^{i})$ $\in$ $\mathcal{A}^{i}$}{
	        \For{$(v_{m}^{i}, v_{n}^{i})$ $\in$ permutation($ia$)}{
	            Initialize $route^{i}$\;
	            $endingNode =  n_{m}^{i}$\;
	            Join(LinkToAnotherCluster\textit{-- Algorithm  \ref{alg:GetNextLink}}) into $cycles$ 
	       }
    }}
    Calculate dist($route_{branch}$) $\forall$ root-to-leaf path in $cycles$\;
    $TSP^* = route_{min(dist)}$\;
    \Output{$TSP^*$}
	\caption{Merging method: this function starts the merging process and calculates the $TSP^*$}
	\label{alg:merging}
\end{algorithm}

\begin{algorithm}[h!]
    \SetKwInOut{Input}{input}
    \SetKwInOut{Output}{output}

	\Input{ $ia(v_{p}^{any}, v_{q}^{any})$,\, $linkedClusters$,\, $\widetilde{\mathcal{TSP}}$, $route^{any}$}
    Get \{$ea(v_{q}^{any},v_{p}^{new})$: $new$ $\notin$ $linkedClusters$, $min(dist(v_{q}^{any},v_{p}^{new}))$\}\;
    Remove($ia$) from $TSP^{any}$\;
    Include($TSP^{any}$) into $route^{any}$\;
    Add($ea$) into $route^{any}$\;
    Join(RecursiveProcedure) into $route^{any}$ \textit{-- Algorithm  \ref{alg:RecursiveSearch}}
    
    \Output{$route^{any}$}
	\caption{LinkToAnotherCluster method: this function builds a bridge to a $TSP^{new}$ and agglutinates the subroute $TSP^{any}$ in the current $route^{any}$ under construction.}
	\label{alg:GetNextLink}
\end{algorithm}

\begin{algorithm}[h!]
    \SetKwInOut{Input}{input}
    \SetKwInOut{Output}{output}

	\Input{$ea = (v_{r}^{any},v_{s}^{new})$,\, $linkedClusters$,\, $\widetilde{\mathcal{TSP}}$}
	Initialize $route^{new}$\;
	Insert($new$, $linkedClusters$)\;
	$ialist$ = Get \{$(v_{s}^{new}, v_{t}^{new}): (v_{s}^{new}, v_{t}^{new}) \in TSP^{new}$\}\;
		\For{$ia(v_{s}^{new}, v_{t}^{new}) \in ialist$}{
		    \If{$new$ is InitialCluster $i$}{
		        lastEdge = $(v_{t}^{new}, endingNode)$\;
		        Add($lastEdge$) into $route^{new}$\;
		        break\;}
		    \Else{
    		    Fork(LinkToAnotherCluster) into $route^{new}$ \textit{-- Algorithm  \ref{alg:GetNextLink}}
	        } 
		}
	\Output{$route^{new}$}
	\caption{RecursiveProcedure method: this function is carried out inside cluster $new$, with the aim of governing the recursive search and generating feasible complete routes.} 
	\label{alg:RecursiveSearch}
\end{algorithm}

Once the merging procedure is completed and $TSP^*$ is built, the QTA calculates its fitness using the classical TSP objective function based on the cost of the traveled distance. After that, the algorithm replaces the current solution only if the new one improves it and checks if the number of accesses to the D-Wave have been reached. In positive case, the QTA returns the best-found solution. If negative, the current solution is modified using the well-known insertion function \cite{osaba2016improved} and re-inserted into the optimization module.

\section{Experimental Setup and Results}\label{sec:exp}

All experiments have been conducted on an Intel Core i7-7600 computer, with 2.90 GHz and a RAM of 16 GB, using one core for computation. Despite the code is adapted for solving the problem with the D-Wave \textit{Advantage\_system1.1} quantum processor with 5436 qubits, and because of the strict limits on the access to the quantum computer, we have used the QBSolv tool offered by D-Wave System in its local variant. The rationale behind this decision is to make a significant and representative experimentation using TSP instances from 14 to 23 nodes. In other case, due to limited access to the computer, the experimentation would be made with extremely small instances, not being able to gain valuable and reliable insights. 

Overall, 7 different instances with $N=16$ to $N=23$ nodes have fed the algorithm. Four of them have been obtained from the widely-used TSPLIB Benchmark: \texttt{Burma14}, \texttt{Ulysses16}, \texttt{Ulysses22} and \texttt{wi29}. The remaining were collected from the reputed Augerat CVRP Benchmark: \texttt{P-n16-k8}, \texttt{P-n19-k2}, and \texttt{P-n23-k8} and reformulated to accomplish the TSP requirements. Main reasons for this benchmark configuration are twofold: their reduced size (which is in concordance with the current state of the art) and their reputation among experts. In order to achieve statistically representative results, each instance has been run 20 times.
Results obtained by the developed technique are compared to those achieved by the QBSolv. On this respect, and embracing the strategies followed by previously published papers \cite{feld2019hybrid,teplukhin2020electronic}, we have used the default configuration of D-Wave’s QBSolv\footnote{At the time of writing, QBSolv’s current version is 0.2.8.}. This configuration includes the \textit{auto\_scale} function, which automatically scales the values of the QUBO matrix to the allowed range of values.

Regarding the parameterization of the developed QTA, the maximum size of clusters has been fixed on 10 nodes due to QC hardware limitations. As termination criteria for the whole algorithm, 40 accesses to the D-Wave computer is the upper bound. Furthermore, the \textit{insertion} successor function guides the search process generating new solutions. Lastly, the size of tabu dictionary is unlimited, aiming at minimizing the number of non-profitable calls to QC resources.

Our QTA has been implemented on the code openly offered by user \textit{mstechly} in GitHub\footnote{https://github.com/BOHRTECHNOLOGY/quantum\_tsp}. Furthermore, the Python implementation of the algorithm proposed in this paper has been made publicly available\footnote{https://github.com/EnekoOsaba/DWAVE4TSP\_Clustering Multitasking} together with the scripts that generate the results next discussed.

Finally, it is convenient to highlight that the principal goal of the designed QTA is to reach promising results, similar to those found by solvers such as QBSolv, while overcoming the main disadvantages of QC and the principal concerns of the scientific community. This way, our main motivations are a) to drastically reduce the number of accesses to the quantum computer, avoiding high economic costs and possible connection issues; and b) to provide the user with the full control of the calls to D-Wave, being able to access the quantum annealer in a more efficient way.

\subsection{Results and Discussion}\label{sec:exp_res}

Table \ref{tab:Results} shows the results (average/standard deviation/best) yielded by both QTA and QBSolv. We also include the percentage difference between the winner and the runner-up with reference to the average results for a comprehensive comparison instance by instance. Furthermore, to fairly assess our key objective regarding the usage of D-Wave's hardware, table \ref{tab:Results} also contains the average accesses needed by QBSolv to reach each solution for each instance.

\begin{table*}[h!]
	\centering
	\caption{Obtained results (average/standard deviation/best) using QTA and QBSolv. Best results have been highlighted in bold.}
	\label{tab:Results}
	\resizebox{1.5\columnwidth}{!}{
		\begin{tabular}{| c | r r r | r r r | r|}
			\hline   
			\multicolumn{1}{|c|}{} & \multicolumn{3}{c|}{QTA} & \multicolumn{4}{c|}{QBSolv}\\
			\hline
			Instance & Avg & Std & Best & Avg & Std & Best & AvAcc\\ 
			\hline
			\texttt{Burma14} & 3355.5 (+0.9\%) & 33.3 & 3323.0 & \textbf{3323.0} & 0.0 & 3323.0 & 259.0\\
			\texttt{Ulysses16} & \textbf{6747.0} & 0.0 & 6747.0 & 6796.5 (+0.7\%) & 35.1 & 6753.0 & 275.3\\
			\texttt{Ulysses22} & \textbf{7158.3} & 99.0 & 7047.0 & 7183.8 (+0.3\%) & 97.2 & 6958.0 &  285.1\\
			\texttt{wi29} & \textbf{30813.3} & 738.4 & 29593.1 & 30892.1 (+0.2\%) & 802.7 & 29292.2 & 474.2\\
			\texttt{P-n16-k8} & 159.5 (+3.3\%) & 2.8 & 156.2 & \textbf{154.4 }& 0.0 & 154.4 & 280.4\\
			\texttt{P-n19-k2} & 177.1 (+3.1\%) & 4.1 & 172.2 & \textbf{171.7} & 0.2 & 171.6 & 305.2\\
			\texttt{P-n23-k8} & 194.0 (+2.2\%) & 5.7 & 185.7 & \textbf{189.7} & 3.1 & 185.3 & 444.8\\
			\hline
		\end{tabular}
	}
\end{table*}

A first look on the results reveals that, in overall, both algorithms perform in a similar way. Our proposed QTA obtains a slight better performance in 3 out of 7 instances (\texttt{Ulysses16}, \texttt{Ulysses22} and \texttt{wi29}) whereas the QBSolv alternative slightly outperforms in the rest of them (\texttt{Burma14}, \texttt{P-n16-k8}, \texttt{P-n19-k2}, and \texttt{P-n23-k8}). At any case, differences are clearly not significant: 0.2\% to 0.7\% in favour of QTA and 0.9\% to 3.3\% when QBSolv emerges as the best alternative. The key differentiating factor between QTA and QBSolv lies in the demand of QC resources made to produce the final solution of each execution. With QTA just accessing the D-Wave's hardware 40 times regardless the execution and the instance to solve, QBSolv increases its demand considerably ranging from an average of 259 accesses for small instances to 474 for the bigger ones.

\begin{figure}[t]
	\centering
	\includegraphics[width=0.7\hsize]{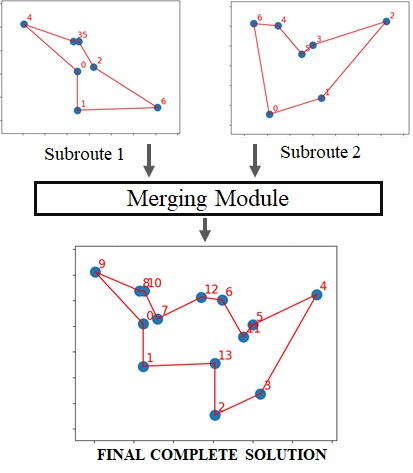}
	\caption{Subroutes and final route for an execution of QTA on Burma14.}
	\label{fig:Burma14}
\end{figure}

\begin{figure}[t]
	\centering
	\includegraphics[width=0.9\hsize]{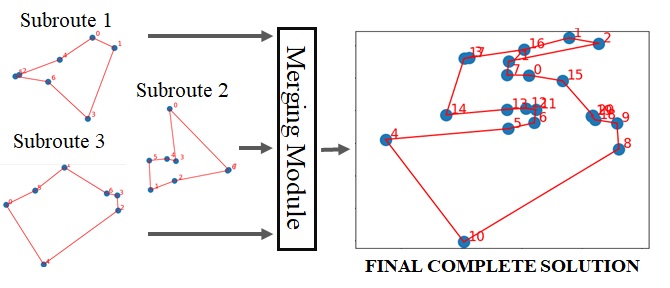}
	\caption{Subroutes and final route for an execution of QTA for Ulysses22.}
	\label{fig:Ulysses22}
\end{figure}

This behavior, along with the competitiveness in terms of results, evinces that QTA brings great advantage by fully meeting the main objective proposed in this research without jeopardizing the performance. In light of these results, we can prudently affirm that QTA is a novel approach which offers promising results for solving partitioning problems while drastically reducing the access to QC resources. Finally, with the intention of complementing the results shown above, we depict in Figure \ref{fig:Burma14} and Figure \ref{fig:Ulysses22} a more detailed example execution of the QTA for the instances \texttt{Burma14} and \texttt{Ulysses22}, respectively.

\section{Conclusions and Future Work}\label{sec:conc}

This work has elaborated on the design, implementation and performance assessment of a complete solving scheme for solving partitioning problems using QC technology. The main pillars of operation are a greater control over the access to quantum resources and a considerable reduction of unnecessary accesses. The principal objective of the developed QTA is to overcome hardware limitations of current QC solvers, offering a framework capable of solving large problem instances. For assessing the quality of our method, we have used the well-known TSP as benchmarking problem. Furthermore, the performance of QTA has been compared to QBSolv on 7 different TSP instances. The experimental outcomes support the preliminary conclusion that QTA is an approach which offers promising results for solving partitioning problems while it drastically reduces non-valuable accesses to QC resources. We plan to devote further efforts in a manifold of research paths rooted on this preliminary study. In the short term, we have planned continue using the TSP as benchmarking problem using larger instances. Additionally, quantum-based schemes for the merging module will be analyzed. In longer term, we will explore the application of the QTA on real-world problems.

\section*{Acknowledgments}

Eneko Osaba, Esther Villar-Rodriguez and Izaskun Oregui would like to thank the Basque Government for its funding support through the EMAITEK and ELKARTEK (through QUANTEK project) programs.

\bibliographystyle{IEEEtran}
\bibliography{IEEEexample}

\end{document}